# Cryogenic Characterization of Ferroelectric Non-volatile Capacitors

Madhav Vadlamani, Dyutimoy Chakraborty, Jianwei Jia, Halid Mulaosmanovic, *Senior Member, IEEE*, Stefan Duenkel, Sven Beyer, Suman Datta, *Fellow, IEEE*, and Shimeng Yu, *Fellow, IEEE*

*Abstract*— Ferroelectric-based capacitive crossbar arrays have been proposed for energy-efficient in-memory computing in the charge domain. They combat the challenges like sneak paths and high static power faced by resistive crossbar arrays but are susceptible to thermal noise limiting the effective number of bits (ENOB) for the weighted sum. A direct way to reduce this thermal noise is by lowering the temperature as thermal noise is proportional to temperature. In this work, we first characterize the non-volatile capacitors (nvCaps) on a foundry 28 nm platform at cryogenic temperatures to evaluate the memory window, ON state retention as a function of temperature down to 77K, and then use the calibrated device models to simulate the capacitive crossbar arrays in SPICE at lower temperatures to demonstrate higher ENOB (~5 bits) for 128x128 multiple-and-accumulate (MAC) operations.

## I. INTRODUCTION

THE non-volatile capacitor (nvCap) leverages C-V asymmetry in small-signal and non-destructive read mode, opening a memory window (MW) with distinct low-capacitance (LCS) and high-capacitance states (HCS) while achieving a high on/off ratio (~25) at 0 V DC voltage. Prior work has demonstrated and analyzed nvCap operation in GlobalFoundries' 28 nm ferroelectric field effect transistor (FeFET) platform at room temperature [1]. The capacitive behavior is governed by polarization-dependent charge modulation in the channel when the source/drain are grounded, while the input signal is applied to the gate. In HCS (see Fig. 1(b)), positive polarization induces an inversion layer that enables full gate-area capacitance contribution through inversion capacitance. Conversely, in LCS, negative polarization results in depletion restricting capacitance to only the gate-to-source/drain overlap region, producing a substantial on/off ratio in capacitance.

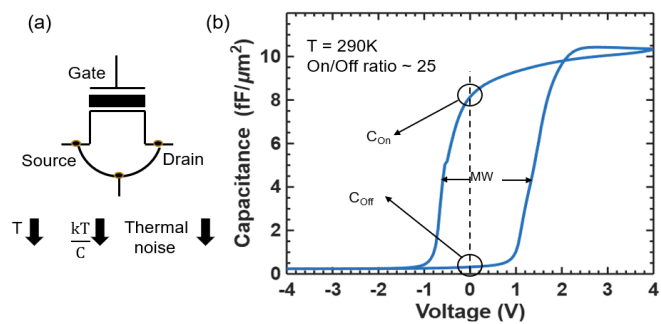

Fig. 1. (a) nvCap, the ferroelectric non-volatile capacitive synapse. (b) C-V characteristics of nvCap with an On/Off ratio of about 25 at 0V and 290K.

Crossbar arrays utilizing these nvCaps have been proposed for parallelizing vector-matrix multiplication (VMM) in the charge domain with only dynamic power, eliminating the static power in the resistive crossbar array [2], [3], [4], [5], [6]. However, the effective number of bits (ENOB) for the partial sum results from multiple-and-accumulate (MAC) operation is fundamentally limited by thermal noise [7], which scales with kT/C. The cryogenic operation of these devices not only suppresses thermal noise to improve ENOB but also makes them particularly attractive for applications requiring low-temperature operation, including: (1) high-performance data center accelerators, (2) quantum computing peripheral circuitry, and (3) electronics for aerospace applications.

In this letter, we characterize nvCaps at cryogenic temperatures with memory window (MW) trends and retention degradation resilience under read stress in Section II. We then discuss the capacitive crossbar array design and ENOB evaluations in Section III, and conclude with challenges and benchmarking in Section IV.

## II. CRYOGENIC CHARACTERIZATION

The C-V characteristics were experimentally measured using a Keithley 4200SCS parameter analyzer integrated with a Janis CCR4 cryogenic probe station. The source/drain terminal was connected to an ammeter, while an AC signal (100 kHz frequency, 50 mV amplitude) was applied to the gate terminal of the GlobalFoundries' 28 nm FeFET devices. Bidirectional quasi-static C-V sweeps (-4V to 4V) were performed on a fresh device at 77K and 290K.

This work was supported by NSF-CCF-2218604. The device fabrication was supported by the European Union within "NextGeneration EU", the Federal Ministry for Economic Affairs and Climate Action (BMWK) in the framework of "Important Project of Common European Interest - Microelectronics and Communication Technologies", under the project name "EUROFOUNDRY".

Madhav Vadlamani, Dyutimoy Chakraborty, Jianwei Jia, Suman Datta, and Shimeng Yu are with the School of Electrical and Computer Engineering, Georgia Institute of Technology, Atlanta, GA 30332 USA (e-mail: shimeng.yu@ece.gatech.edu).

Halid Mulaosmanovic, Stefan Duenkel, and Sven Beyer are with GlobalFoundries Fab1 LLC & Company KG, 01109 Dresden, Germany.

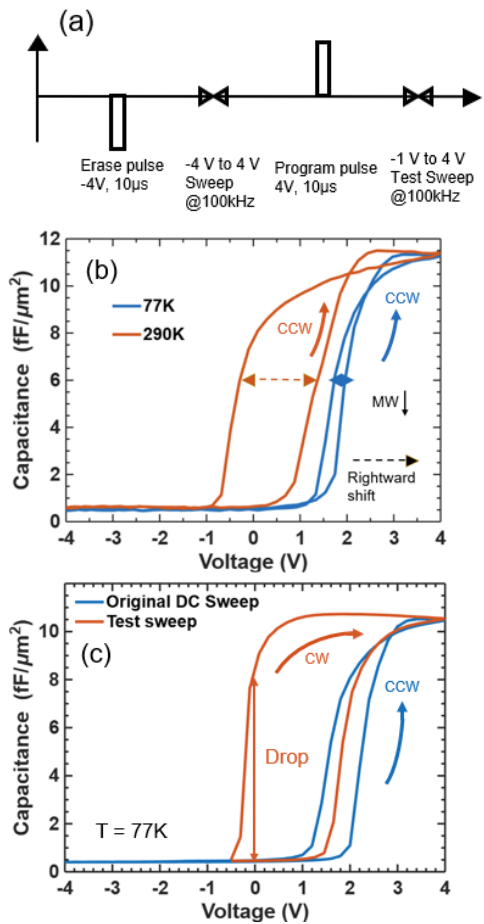

Fig. 2. (a) The pulse scheme used in this experiment. (b) MW shrinkage and rightward shift of the C-V curve during the quasi-static sweep at cryogenic temperatures. (CW: clockwise; CCW: counterclockwise.) (c) Drop in On state capacitance after performing the test sweep on a separate device confirming carrier trapping to be the mechanism behind the MW shrinkage.

As shown in Fig. 2(b), the MW shrinks at 77K (reducing from approximately 2 V to 0.5 V), accompanied by a rightward shift in the C-V curve. This shift is attributed to the reduction in intrinsic carrier density at cryogenic temperatures, which increases the voltage required for channel inversion.

To investigate whether MW narrowing stems from carrier trapping in gate-stack defects [8], a test pulse was conducted on a separate device at 77K, as shown in Fig. 2(a), followed by a bidirectional C-V sweep (−1V to 4V) at the same temperature. Fig. 2(c) reveals a large MW still exists at 77K after the pulse programming, a significant drop in capacitance during the backward sweep compared to the forward sweep in a clockwise direction, suggesting that the enhanced carrier trapping occurs at lower temperatures in the quasi-DC C-V sweep. Crucially, the voltage range was constrained to prevent domain flipping during the test sweep [8], ensuring that the observed hysteresis originates from carrier trapping effects.

We designed a pulse scheme (Fig. 3(a)) to measure the on-state capacitance before charge trapping occurs. The sequence begins with a -4 V, 10 µs erase pulse, followed by a unidirectional -1 V to +1 V DC read C-V sweep. Next, a 4 V, 10 µs program pulse is applied, followed by another unidirectional -1 V to 1 V DC read C-V sweep. As shown in Fig. 3(b), this scheme reveals a large capacitive memory window with an on/off ratio of ~10.

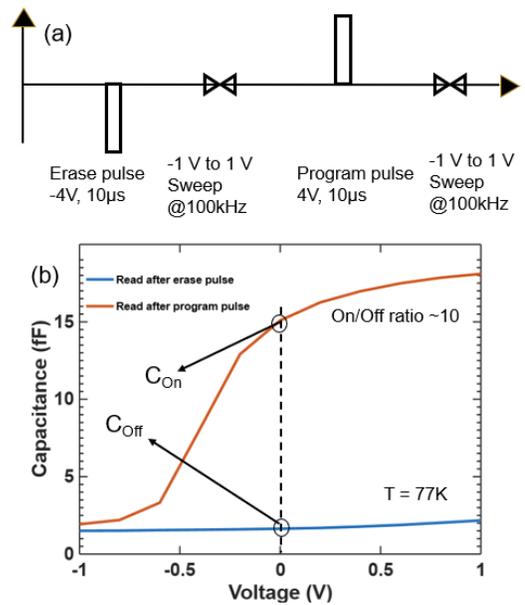

Fig. 3. (a) The pulse scheme used in the experiment. (b) HCS observed being programmed with a 4V, 10us pulse.

Since VMM in the inference mode requires repeated read operations under DC bias, we evaluated device retention using 0.1 V DC read stress measurements. The test protocol (Fig. 4(a)) involves initial device programming followed by a bidirectional DC read C-V sweep (-1 V to +1 V) under constant 0.1 V DC stress. The results (Fig. 4(b)) demonstrate significantly reduced on-state capacitance decay at 77 K compared to 290 K. This improvement aligns with prior findings that domain-wall depinning is thermally activated,

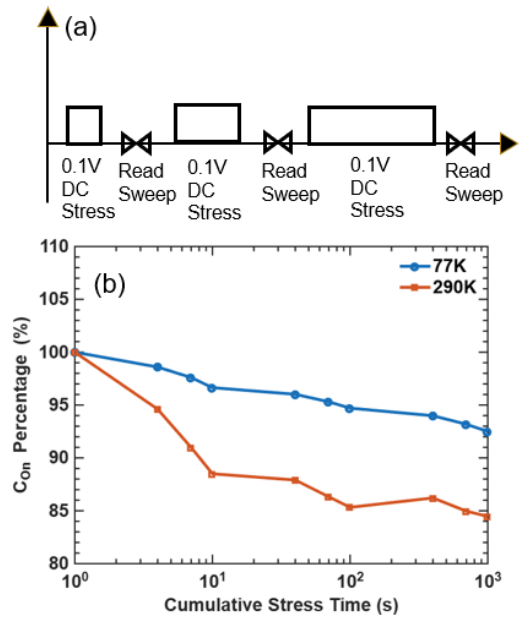

Fig. 4. (a) The pulse scheme used: 0.1V DC stress applied with read sweeps performed periodically. (b) % decay in HCS when subjected to 0.1V DC Stress.

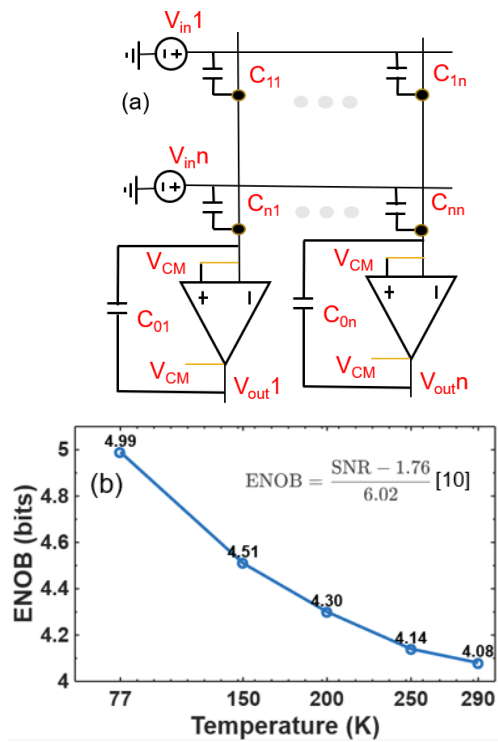

Fig. 5.(a) The schematic of the crossbar array. (b) The ENOB versus temperature, showing ~5 bits at 77 K and ~4 bits at 290 K.

leading to depolarization and enhanced polarization retention at cryogenic temperatures.[9]

## III. Capacitive Crossbar Array

The capacitive crossbar array operates through two phases [2], [3]: (1) a charging phase where wordline (WL) voltages activate the array capacitances, and (2) a discharge phase where the accumulated charge is transferred to a reference capacitor, effectively implementing MAC operations.

Having demonstrated the superior reliability of nvCaps at cryogenic temperatures, we evaluated their performance in a 128×128 crossbar array using SPICE simulations. The simulations incorporated a foundry PDK for peripheral circuitry, with temperature conditions set at various temperatures. The circuit topology follows previous implementations [2], [3], [7] as shown in Fig. 5(a), where weight elements employ compact models of the characterized nvCaps and peripheral circuits incorporate 9T telescopic operational amplifiers.

The effective number of bits (ENOB) is fundamentally determined by the signal-to-noise ratio (SNR) [10], which is inversely proportional to kT/C thermal noise. While this noise originates from both array capacitances and operational amplifiers, we neglected the amplifier's contribution as it is significantly lower than the array-induced noise, per our estimation.

Fig. 5(b) illustrates the temperature dependence of the ENOB in the system. As temperature decreases from 290 K to 77 K, the ENOB exhibits a monotonic improvement from approximately ~4 bits to ~5 bits, reflecting enhanced precision under cryogenic conditions. These results align with theoretical expectations of reduced thermal noise at cryogenic temperatures, confirming the device's suitability for low-temperature applications.

## IV. Conclusion

This work demonstrates that non-volatile capacitors (nvCaps) exhibit enhanced performance as weight elements for in-memory computing applications at cryogenic temperatures. Our experimental characterization reveals improved retention and reduced thermal noise at 77 K, directly improving MAC accuracy. These results are corroborated by SPICE simulations of a 128×128 crossbar array, where the ENOB improves by ~1 bit (from 4 to 5) under cryogenic (77 K) operation compared to room temperature (290K).

## V. Acknowledgements

The authors gratefully acknowledge Junmo Lee and Omkar Phadke for their valuable technical discussions and insightful suggestions that contributed to this work.